\documentclass[12pt]{article}
\usepackage[margin=1in, paperwidth=8.5in, paperheight=11in]{geometry}
\usepackage{amsmath}
\usepackage{graphicx}%
\usepackage{amsfonts}%
\usepackage{amssymb}
\usepackage{color}
\usepackage{float,subfigure}
\usepackage[round]{natbib}                  
\usepackage{cite}
\usepackage{setspace}
\usepackage{comment}
\usepackage{url}
\usepackage{enumerate}
\usepackage{verbatim}
\usepackage{hyperref}
\hypersetup{pdfborder = {0 0 0}} 
\hypersetup{pdfpagemode=UseNone}

\usepackage{caption} 
\usepackage{float}
\usepackage{booktabs}
\usepackage{longtable}

\begin{document}
\thispagestyle{empty}
\setcounter{page}{0}

\begin{center}
\singlespacing\textbf{\Large Evaluating Racialized Economic Segregation in the Presence of Spatial Autocorrelation}
\singlespacing\textbf{Yang Xu, and Loni Philip Tabb}\\
Department of Epidemiology and Biostatistics, Drexel University,\\ Philadelphia, PA 19104
\end{center}

\textsc{Abstract.} 

Research on residential segregation has been active since the 1950s and originated in a desire to quantify the level of racial/ethnic segregation in the United States. The Index of Concentration at the Extremes (ICE), an operationalization of racialized economic segregation that simultaneously captures spatial, racial, and income polarization, has been a popular topic in public health research, with a particular focus on social epidemiology. However, the construction of the ICE metric usually ignores the spatial autocorrelation that may be present in the data, and it is usually presented without indicating its degree of statistical and spatial uncertainty. To address these issues, we propose reformulating the ICE metric using Bayesian modeling methodologies. We use a simulation study to evaluate the performance of each method by considering various segregation scenarios. The application is based on racialized economic segregation in Georgia, and the proposed modeling approach will help determine whether racialized economic segregation has changed over two non-overlapping time points.


\newpage

\section{Introduction}

The debate on the nature of residential segregation and how to measure it has been an active research topic since the 1950s, and it originated in a desire to quantify the level of race/ethnicity segregation in the United States \citep{massey1987trends}. Segregation is defined as the physical separation of two or more groups into different neighborhoods, which was imposed by legislation, supported by numerous economic institutions, enforced by the judicial system, and legitimized by the White supremacist ideology that was endorsed by churches and other cultural groups \citep{massey1987trends, jaynes1989common, massey1993american}.
An increasing amount of public health literature shows that residential segregation is a fundamental cause of racial disparities in health and links patterns of residential segregation to various health status, including mortality, morbidity, health behavior issues, and mental illness \citep{larrabee2022racialized}. In particular, segregation is a major driver of inequalities between Blacks and Whites since it significantly impacts the socioeconomic status of Blacks at individual, household, neighborhood, and community levels \citep{williams2001racial}. 

Over the years, a variety of indexes have been proposed to measure residential segregation from different perspectives. Based on the well known study by \citet{massey1987trends}, the segregation indexes can be classify into five key dimensions: (1) measures of evenness - comparing the spatial distribution of different groups among units in a metropolitan area, including the dissimilarity index and the Gini coefficient; (2) measures of exposure - quantifies the degree of potential contact or interaction between members of the minority and the majority group, including interaction and isolation indexes; (3) measures of concentration - refers to the proportional quantity of metropolitan space occupied by a minority group, including delta, absolute concentration and relative concentration indexes; (4) measures of centralization - measures the degree to which a group is spatially located near the center of an urban area, including absolute and relative measures; and (5) measures of clustering - quantifies the extent to which areal units inhabited by minority groups adjoin or cluster in space, including absolute clustering, index of spatial proximity and relative clustering. Research built on these indexes has been considerably varied in the past, including having multiple racial/minority sub-groups \citep{reardon2002measures, reardon2004measures}, addressing the modifiable areal unit problem \citep{wong2003spatial,simpson2007ghettos}, critiquing its formulae \citep{lee2015bayesian}, and extending its usage on public health monitoring \citep{krieger2016public}.

Residential segregation has consistently been shown to shape racial differences in socioeconomic status, particularly among Black populations who are highly segregated and have little access to education and employment opportunities, resulting in racial income inequality \citep{popescu2018racial}. However, the above indexes have been used to characterize one aspect of residential segregation - racial segregation, whereas information on income segregation has been largely ignored. There is a greater need for an index that can simultaneously measure racial and income segregation. Therefore, in this study, we consider evaluating the Index of Concentration at the Extremes (ICE), a metric that was first developed by Massey,
to reveal the extent to which an area's residents are concentrated into groups at the extremes of deprivation and privilege, and has been extensively studied in the social sciences and etiological public health investigations. Krieger and colleagues recently extended the ICE metric to simultaneously measure race/ethnicity and income without changing its original formulation, which has been considered an operationalization of racialized economic segregation (ICE$_{race+income}$) \citep{krieger2016metrics,krieger2016public}. Although the new ICE metric has been extensively employed in social epidemiology studies, and the objectives of these studies have typically focused on evaluating the impact of ICE on a variety of health and social outcomes, in this study, we take an alternative perspective by considering evaluating the ICE metric itself based on two issues that have not been addressing in the ICE related research. Our study is motivated by Lee's research, in which he addressed two issues that have been overlooked in the literature on residential segregation: (1) measure of uncertainty and (2) spatial correlation, where the application of his study was based on reconstructing the Dissimilarity Index for religious segregation in Northern Ireland. \citep{lee2015bayesian}. These two statistical issues have also occurred when computing the ICE metric. First and foremost, the ICE metric is more of a descriptive summary statistic that usually presents without indicating its degree of statistical uncertainty. Accounting for the uncertainty in the segregation indexes, as addressed by multiple studies, could assist researchers in determining whether the observed differences in the indices represent a meaningful change or are the result of sampling variation/errors. Second, we argue that the computation of the ICE metric typically neglects the spatial correlation that may present in the data. However, since the ICE is essentially a function of three distinct populations, the quantification of spatial and statistical uncertainty remains a challenge; thus, in this study, we propose a reformulation of the ICE metric, which enable us to quantify the uncertainty and the spatial correlation in the data, without losing the original interpretation of the ICE metric. We propose three different approaches to addressing these two issues, and we use simulation studies to compare the performance of each approach objectively. We do not attempt to challenge the existing literature that address the relationship between the ICE and various health outcomes, rather, we want to provide a unique perspective for public health researchers to consider the critical components behind this metric.



The structure of our paper is as follows: Section 2 describes the data and motivated study. Section 3 describes the methodology, including the reformulation of the ICE measure and the three statistical approaches proposed for computing the new ICE measure. In section 4, we conduct a simulation study to compare the properties of the three approaches in an objective manner. The simulation results are presented in section 5, while the results of the real data analysis are presented in section 6. Section 7 concludes the paper by discussing the significant findings of the study and future ICE related research.


\section{Data and Motivation}

The study is motivated by the racialized economic segregation in Georgia, a state with a large African American population in the southeastern US. 
According to the most recent data from the United States Census Bureau, 51.9 percent of the population is White in Georgia, down from 59.7 percent in 2010; where the Black population has increased slightly over the last decades, from 31.5 percent to 33 percent. These changes in the distributions of these racial groups could have an impact on the distribution of racialized economic segregation in Georgia over time. As racialized economic segregation, by way of ICE, has been proved as an essential part of social determinants of health, this change in ICE may also potentially affect the health of the population in Georgia. Therefore, in this study, we aim to assess the racialized economic segregation at the county level in Georgia at two non-overlapping time points (2005-2009 vs. 2016-2020), to evaluate whether there are substantial changes in racialized economic segregation. 

\begin{figure}[ht]
   \centering 
   \includegraphics[height=10cm, width=14cm]{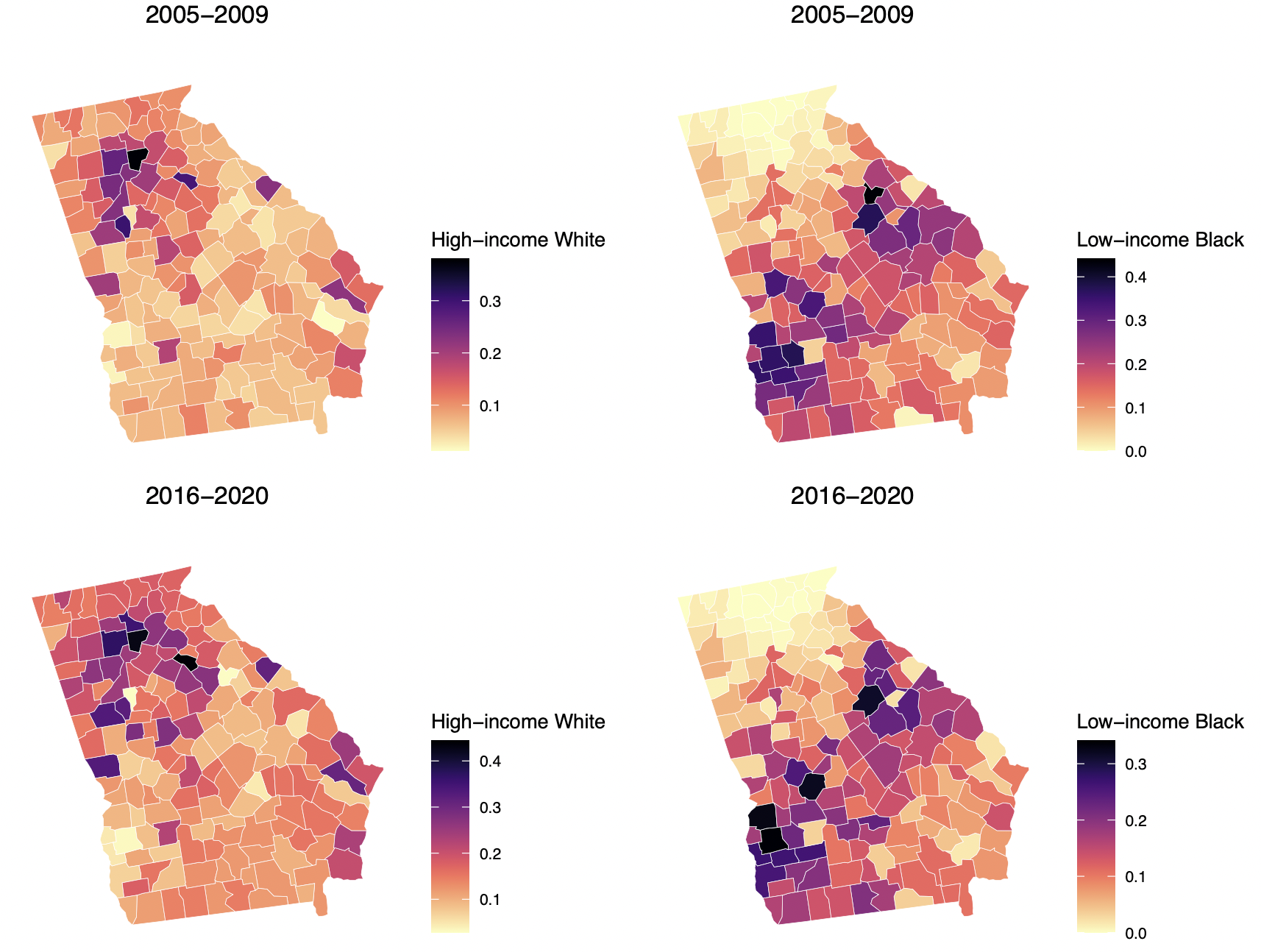}
   \caption{Maps of the raw proportions of high-income White and low-income Black residents in each county for 2005-2009 (top row) and 2016-2020 (bottom row) 5-year American Community Survey estimates. }
   \label{fig:ga_prop}
\end{figure}

Data for both years are obtained at the Georgia county level from the American Community Survey 5-year estimated data, where includes the number of non-Hispanic White residents whose income is in the 80th percentile, number of non-Hispanic Black residents whose income is in the 20th percentile, and the total number of residents whose income is known.  The raw proportions of high-income White and low-income Black residents in each county in 2009 and 2020 are displayed in Figure \ref{fig:ga_prop}. The top two maps illustrate each proportion group in the 5-year estimate for 2005-2009, while the bottom maps represent each group in the 2016-2020 estimate. These maps clearly demonstrate that each proportion group exhibits some degree of spatial correlation, as geographically adjacent counties tend to have similar proportion values. This observation is supported by a Moran's I test, which reveals that each group has statistically significant I statistics of 0.412 (high-income White, 2009), 0.597 (low-income Black, 2009), 0.415 (high-income White, 2020), and 0.535 (low-income Black, 2020). These findings further suggest the importance of accounting for spatial correlation and data uncertainty when estimating proportion groups at the county level. On the other hand, Figure 1 demonstrates that the spatial correlation of high-income White and low-income Black groups is localized since some geographically adjacent counties have remarkably different proportions. This may violate the assumption of spatial correlation, as adjacent areas should have similar values. Furthermore, these step changes between geographic units are known as boundaries in the disease mapping field, and previous studies suggest that global smoothing models might not be able to capture this type of spatial structure in the data \citep{lee2012boundary, lee2013locally}. As a result, in addition to utilizing global spatial models to control for spatial correlation in data, this paper proposes a localized smooth model to capture the complex spatial structure described above, where the proposed model extends previous research that aimed to also examine the uncertainty and spatial correlation when utilizing these types of residential segregation measures.

The Index of Concentration at the Extremes (ICE), a novel measure that summarizes both racial and income segregation in a single metric, has been extensively studied and applied in population health research, with a specific focus on social epidemiology. The ICE, particularly the ICE$_{race+income}$, an operational measure of racialized economic segregation, has been shown to impact different health outcomes such as adverse birth/pregnancy outcomes \citep{krieger2018using}, cancer outcomes \citep{krieger2018cancer}, premature mortality \citep{krieger2016monitoring}, COVID-19 \citep{chen2021revealing}, and cardiovascular health \citep{tabb2022spatially}. Mathematically, the ICE$_{race+income}$ (hereafter ICE) is defined as $\text{ICE}_{i}=\frac{A_{i}-P_{i}}{T_i}$, where $A_{i}$ is the number of affluent persons (greater than or equal to 80th income percentile) in neighbourhood $i$, $P_{i}$ is the number of poor persons (less than or equal to 20th income percentile) in neighborhood $i$ and $T_i$ is the total population for whom income level is known in neighborhood $i$. The value of ICE lies in the interval $[-1,1]$ where -1 indicates the most deprived group and 1 is the most privileged group. While the ICE metric can be used to compare various race/ethnicity with income groups, we focus on Black-White comparisons given the motivated study in Georgia.

\vspace{5mm}

\section{Method}
\subsection{Reformulating the racialized economic segregation measure}

In this study, we propose a reformulation of the ICE measure by accounting for the spatial correlation and uncertainty that we hypothesized to be present in the data. 
Suppose we reformulate the ICE equation as ICE$_i=\frac{A_i}{T_i}-\frac{P_i}{T_i}$, that is, the ICE measure is a function of two sample proportions, where the first component $\frac{A_i}{T_i}$ represents the sample proportion of high-income White residents and the second component $\frac{P_i}{T_i}$ represents the sample proportion of low-income Black residents. In this way, we demonstrate a distinct view of thinking about the ICE measure without changing its value, i.e., the new formulation of ICE is still between -1 and 1, with the exact same interpretation as the original ICE measure. Each neighborhood $i$ will still have an ICE measure, indicating whether an area's residents are concentrated into groups at the extremes of deprivation or privilege. This approach would allow us to quantify and estimate each component of the formula while accounting for spatial correlation and uncertainty. Since the two proportions represent distinct population groups, the level of uncertainty and spatial correlation between them will likely differ. As a result, we propose estimating each group separately and then combining them to obtain an estimated posterior distribution of the ICE measure.


\subsection{Modelling approaches}
The statistical approaches described below will represent the two extremes being compared when considering the ICE measure as an operationalization of racialized economic segregation; such that, (1) represents high-income White and (2) low-income Black. Specifically, consider $Y_i^{(1)}$ is the observed number of high-income White residents in each county $(i=1,...,n)$, $p_i^{(1)}$ is the unknown proportion of high-income White in each county. Similarity, $Y_i^{(2)}$ is the observed number of low-income Black residents in each county, $p_i^{(2)}$ is the unknown proportion of low-income Black in each county. Based on the formulation of ICE, $N_i$ represents the total number of resident with known income in each county, that is, the two proportion groups share the same $N_i$.

We propose to use three different approaches: (1) naive approach (bootstrap), (2) globally smooth models, and (3) locally smooth models to estimate the ICE metric based on the proposed reformulation.
\subsubsection*{Approach 1: Bootstrapping}
The first approach that we consider is based on the simple method of moments estimator, and the $95\%$ confidence interval is computed using a bootstrapping approach. We create 10,000 pseudo data sets, by resampling the data point of each groups with replacement. Among each resampled data set, an estimate ICE measure is computed, and a $95\%$ confidence interval is constructed by calculating the 2.5 and 97.5 percentiles of the set of ICE values. 
\vspace{2mm}
\begin{gather*}
ICE_{i}=\hat p_i^{(1)}-\hat p_i^{(2)}\\
\hat p_i^{(1)}=\frac {Y_i^{(1)}}{N_i} \qquad 
\hat p_i^{(2)}=\frac {Y_i^{(2)}}{N_i}
\end{gather*}

\subsubsection*{Approach 2: Globally smooth models}
The globally smooth model that we consider in this paper is based on the binomial generalized linear mixed effect model, where random effects are assumed spatially correlated \citep{banerjee2003hierarchical}. Specifically, consider $\boldsymbol v=(v_1,...,v_n)$ to be the set of random effects in the globally smooth models, and can be described by different conditional autoregressive (CAR) priors. An important part of the CAR priors is to define a binary $n\times n$ adjacency matrix $\boldsymbol{W}=(w_{ij})$, which captures the spatial proximity structure. The common approach is to assume areas $(i\sim j)$ are neighbors, $i.e. w_{ij}=1$ if and only if $i$ and $j$ share a border, otherwise $w_{ij}=0$. The spatial CAR priors that were considered for globally smooth models are: Besag-York-Mollié model (BYM), Intrinsic CAR model (ICAR), and Leroux model (Leroux). The models that are utilized here are consistent with existing research \citep{goldstein2003modelling, lee2015bayesian}, and the full models based on different spatial priors are described below: 

\paragraph{(1) Besag-York-Mollié model} \mbox{}\\
The first model we consider is the Besag-York-Mollié model \citep{besag1991bayesian}, one of the most well-known models for areal disease mapping. The full model is given by: 
\vspace{2mm}
\begin{gather*}
ICE_{i}=p_i^{(1)}-p_i^{(2)}\\
Y_i^{(1)}\sim \text{Binomial} (N_i,p_i^{(1)})\qquad
Y_i^{(2)}\sim \text{Binomial} (N_i,p_i^{(2)})\\
ln\left(\frac{p_i^{(1)}}{1-p_i^{(1)}}\right)=\beta_0^{(1)}+v_i^{(1)}+u_i^{(1)}\qquad
ln\left(\frac{p_i^{(2)}}{1-p_i^{(2)}}\right)=\beta_0^{(2)}+v_i^{(2)}+u_i^{(2)}\\
p_i^{(1)}= \frac{exp\left(\beta_0^{(1)}+v_i^{(1)}+u_i^{(1)}\right)}{1+exp\left(\beta_0^{(1)}+v_i^{(1)}+u_i^{(1)}\right)}\qquad
p_i^{(2)}= \frac{exp\left(\beta_0^{(2)}+v_i^{(2)}+u_i^{(2)}\right)}{1+exp\left(\beta_0^{(2)}+v_i^{(2)}+u_i^{(2)}\right)}
\end{gather*}
where $v_i$ and $\mu_i$ represent the spatial structured and unstructured terms for county $i$, respectively. The spatial structured term is specified as $\left(v_i \mid\boldsymbol{v_{j}}, i\neq j,\sigma^2_{v} \right)\sim N      \left(\frac{\sum_{j}v_{j}w_{ij}}{\sum_{j}w_{ij}},\frac{\sigma^2_{v}}{\sum_{j}w_{ij}}  \right)$, and the unstructured term $u_i$ is assigned a Normal distribution $u_i\sim N(0,\sigma^2_{u})$, where the prior for each variance $\sigma^2_{v}$ and $\sigma^2_{u}$ is assigned by a vaguely informative prior, namely inverse-gamma (1,0.01). The intercept $\beta$ is assigned a flat prior. Based on the modeling setup, each proportion group can be achieved by using the inverse logit transformation from the posterior samples. Thus a new posterior estimate of the ICE metric is computed. Similar modelling setup is applied to the rest of globally smooth models as well as the locally smooth models. We present a sensitivity study and show that the posterior estimates are not impacted by changing the priors.

\paragraph{(2) Intrinsic CAR model} 
\mbox{}\\
The Intrinsic CAR model was the first CAR prior to be proposed \citep{besag1974spatial}, and serves as the foundation for other priors such as the BYM. The ICAR model has a single random effect to capture the spatial autocorrelation in the data. The model is given by:
\vspace{1mm}
\begin{gather*}
ICE_{i}= p_i^{(1)}- p_i^{(2)}\\
Y_i^{(1)}\sim \text{Binomial} (N_i^{(1)},p_i^{(1)})\qquad
Y_i^{(2)}\sim \text{Binomial} (N_i^{(2)},p_i^{(2)})\\
ln\left(\frac{p_i^{(1)}}{1-p_i^{(1)}}\right)=\beta_0^{(1)}+v_i^{(1)}\qquad
ln\left(\frac{p_i^{(2)}}{1-p_i^{(2)}}\right)=\beta_0^{(2)}+v_i^{(2)}\\
p_i^{(1)}= \frac{exp\left(\beta_0^{(1)}+v_i^{(1)}\right)}{1+exp\left(\beta_0^{(1)}+v_i^{(1)}\right)}\qquad
p_i^{(2)}= \frac{exp\left(\beta_0^{(2)}+v_i^{(2)}\right)}{1+exp\left(\beta_0^{(2)}+v_i^{(2)}\right)}
\end{gather*}

where $v_i$ is assigned an intrinsic CAR prior, which is specified as $\left(v_i \mid\boldsymbol{v_{j}}, i\neq j,\sigma^2_{v} \right)\sim N      \left(\frac{\sum_{j}v_{j}w_{ij}}{\sum_{j}w_{ij}},\frac{\sigma^2_{v}}{\sum_{j}w_{ij}}  \right)$, and the prior for the variance $\sigma^2_{v}$ is assigned by a inverse-gamma (1,0.01). The intercept $\beta$ is assigned the same flat prior as the BYM prior shown previously.

\paragraph{(3) Leroux model} 
\mbox{}\\
The Leroux model, proposed by Leroux, is another model that has been widely used in spatial contexts \citep{leroux2000estimation}. The Leroux model is a generalization of the independent and ICAR models since it contains an extra parameter serving as the weight to control the degree of spatial correlation in the data. The model is given by: 
\vspace{3mm}
\begin{gather*}
ICE_{i}=\hat p_i^{(1)}-\hat p_i^{(2)}\\
Y_i^{(1)}\sim \text{Binomial} (N_i,p_i^{(1)})\qquad
Y_i^{(2)}\sim \text{Binomial} (N_i,p_i^{(2)})\\
ln\left(\frac{p_i^{(1)}}{1-p_i^{(1)}}\right)=\beta_0^{(1)}+v_i^{(1)}\qquad
ln\left(\frac{p_i^{(2)}}{1-p_i^{(2)}}\right)=\beta_0^{(2)}+v_i^{(2)}\\
p_i^{(1)}= \frac{exp\left(\beta_0^{(1)}+v_i^{(1)}\right)}{1+exp\left(\beta_0^{(1)}+v_i^{(1)}\right)}\qquad
p_i^{(2)}= \frac{exp\left(\beta_0^{(2)}+v_i^{(2)}\right)}{1+exp\left(\beta_0^{(2)}+v_i^{(2)}\right)}
\end{gather*}
where $v_i$ is assigned a Leroux prior, which specified as $\left(v_i \mid\boldsymbol{v_{j}}, i\neq j,\sigma^2_{v} \right)\sim N \left(\frac{\rho\sum_{j}v_{j}w_{ij}}{\rho\sum_{j}w_{ij}+1-\rho}, \frac{\sigma^2_{v}}{\rho\sum_{j}w_{ij}+1-\rho} \right)$. The $\rho$ is the extra parameter that controls the spatial correlation structure and ranges from 0 to 1. When $\rho=0$, this prior is an independent model with a constant mean and variance, and when $\rho=1$, it is an intrinsic CAR model. The prior for the variance and other parameters are the same as in the previous models, and a sensitivity analysis is performed.

\subsubsection*{Approach 3: Locally smooth models}

The concept of locally smooth models that we propose here is similar to Lee's modeling framework \citep{lee2012boundary, lee2013locally, lee2015bayesian} in that it allows the neighborhood random effects to differ substantially across geographic space—as we empirically observed by the proportions of low-income Black residents at the county level across Georgia (Figure 1). We will first review Lee's model and then propose an alternate formulation based on Lee's methodology.

In Lee's study, the proposed model has the following structure:
\begin{gather*}
Y_i\sim \text{Binomial}(N_i,p_i)\\
ln\left(\frac{p_i}{1-p_i}\right)=\beta_{z_{i}}+v_i\\
Z_i\sim\text{Multinomial} \left(1;1/G,...,1/G\right)\\
v_{i}\sim N(\boldsymbol{0}, \sigma^2Q(\rho,W)^{-1})\\
\beta_j\sim \text{Uniform}(\beta_{j-1},\beta_{j+1})\hspace{2mm} \text{for}\hspace{2mm} j=1,...,q\\
\sigma^2\sim \text{Inverse-Gamma}(0.001,0.001)\\
\rho \sim \text{Uniform}(0,1)
\end{gather*}
This model combines the random effects with a piecewise constant intercept term to allow sudden jumps in the spatial surface between adjacent areal units. Specifically, the intercepts $\boldsymbol{\beta}$ is constrained so that $\beta_1<\beta_2...<\beta_q$, where $\beta_0=-\infty$ and $\beta_{q+1}=\infty$. The variable $Z_i$ is defined as a set of cluster indicators, with each indicator $Z_i \in \{1,...,q\}$, which $Z_i$ will assign to each study area and determine which groups those areas belong to. A multinomial distribution with equal probabilities is assigned to $Z_i$, and the cluster $q$ must be defined by prior knowledge; typically, comparison between different values of $q$ is required in sensitivity analyses. Additionally, since each area can only have one intercept term, the multinomial distribution above has a number $\boldsymbol{1}$, indicates there is one trial  in the multinomial distribution. The piecewise intercept component does not account for any spatial correlation in the data; thus, a random effect $v_i$ is introduced in the model to account for spatial correlation by employing the Leroux CAR prior. 

\subsubsection*{Approach 4: Alternative formulation of locally smooth models
}
In this paper, we propose an alternative locally smooth modeling approach that serves the same purpose as Lee's locally smooth model but with different prior distributional assumptions. The model is given by:
\vspace{2mm}
\begin{gather*}
ICE_{i}=\hat p_i^{(1)}-\hat p_i^{(2)}\\
Y_i^{(1)}\sim \text{Binomial} (N_i,p_i^{(1)})\qquad
Y_i^{(2)}\sim \text{Binomial} (N_i,p_i^{(2)})\\
ln\left(\frac{p_i^{(1)}}{1-p_i^{(1)}}\right)=\beta_{z_{i}}^{(1)}+\phi_i^{(1)}\qquad
ln\left(\frac{p_i^{(2)}}{1-p_i^{(2)}}\right)=\beta_{z_{i}}^{(2)}+\phi_i^{(2)}\\
p_i^{(1)}= \frac{exp\left(\beta_{z_{i}}^{(1)}+\phi_i^{(1)}\right)}{1+exp\left(\beta_{z_{i}}^{(1)}+\phi_i^{(1)}\right)}\qquad
p_i^{(2)}= \frac{exp\left(\beta_{z_{i}}^{(2)}+\phi_i^{(2)}\right)}{1+exp\left(\beta_{z_{i}}^{(2)}+\phi_i^{(2)}\right)}
\end{gather*}

where $\beta_j\sim \text{Uniform}(-\infty,\infty)$, for $j=1,...,q$. The $q$ area intercepts $\boldsymbol{\beta}$ is constrained such as $\beta_1 < \beta_2<...<\beta_q$, potentially prevents the label switching issue in the models. The vector $\boldsymbol{Z}=(Z_1,...,Z_k)$ contains a set of cluster indicators, where $Z_i \in \{1,...,q\}$. Instead of a multinomial prior, we propose to use a categorical prior for the indicator $Z_i$, such that $Z_i\sim Cat(p_1,...,p_q)$, with each indicator has equal probabilities. Recall that in Lee's study, $Z_i$ was assigned a multinomial prior with a single trial, as each study areal unit can only contain a single intercept component. In this case, the categorical distribution is actually a special case of multinomial with only one trial. Similar to Lee's model, the piecewise constant intercept that we propose is non-spatial, since we include an additional random effect in the model to capture the spatial correlation in the data. The random effect $\phi_i$ is assigned a BYM prior, where $\phi_i=v_i+u_i$, which represents the spatial structure and unstructured terms, respectively. The spatial structured term is specified as $\left(v_i \mid\boldsymbol{v_{j}}, i\neq j,\sigma^2_{v} \right)\sim N\left(\frac{\sum_{j}v_{j}w_{ij}}{\sum_{j}w_{ij}},\frac{\sigma^2_{v}}{\sum_{j}w_{ij}}  \right)$, and the unstructured term $u_i$ is assigned a Normal distribution $u_i\sim N(0,\sigma^2_{u})$, each variance $\sigma^2_{v}$ and $\sigma^2_{u}$ is assigned by a vaguely informative prior, namely inverse-gamma (1,0.01). An essential part of this model is that it depends on $q$, the number of clusters in the piecewise constant intercept term, which is expected to be known in advance \citep{lee2015bayesian}. However, it is challenging to determine the optimal number of clusters for a given study region. Thus, we fit models with different values of clusters $q=1,...,3$, where $q=1$ is the global smoothing model with a common intercept over the study region. We use the Watanabe–Akaike information criterion (WAIC) as a model selection criterion to determine the optimal value of $q$ based on the study region, and the posterior estimate of the ICE metric can be computed using the aforementioned modeling setup \citep{watanabe2010asymptotic}. 

Bayesian inference for the globally and the locally smooth models are based on Markov chain Monte Carlo (MCMC) simulation. The sampler for these models were run for 50,000 iterations with with the first 20,000 iterations discarded as burn-in. All models are fitted in R, via R-NIMBLE \citep{de2017programming} packages.

\section{Simulation study}
To examine the estimation properties of the ICE measure, we turn to simulations that capture various segregation scenarios to compare the statistical approaches described previously. 
We generate simulated racialized economic segregation data for the set of $n=159$ counties that comprise Georgia state, which is the study region for the motivating application. The proportions of high-income White $(p_1^{(1)},...,p_n^{(1)})$, and low-income Black $(p_1^{(2)},...,p_n^{(2)})$ residents are generated as below. The total population sizes for each region $(N_1,...N_n)$ are assumed known, however, are varied to determine their impact on model performance. Binomial sampling is used to generate the simulated number of high-income White $(Y_1^{(1)},...,Y_n^{(1)})$ and number of low-income Black $(Y_1^{(2)},...,Y_n^{(2)})$ residents in a region. We generate one hundred simulated data sets under two cases, with 4 different scenarios. The difference between cases 1 and 2 is that case 1 assumes that the two racial groups have a similar proportion in each county, whereas case 2 assumes that one group has a larger proportion in the majority of the county. Details of the simulation study are summarized below.

\begin{flushleft}
\textbf{Case 1}\\
\textbf{Low Racialized Economic Segregation (balance between deprived and privileged) with spatial variation:} We assume that the true low-income Black proportions are similar to high-income White proportions in a region.
\end{flushleft}
\textbf{Scenario 1:} Similar proportions of low-income Black and high-income White residents with low spatial correlation. A multivariate Gaussian distribution is used to generate the logit transformation of the proportions with means of -1.72 $(p_1=p_2=0.15)$, with a variance of 0.2. Spatial correlation is based on a proper CAR model with $\rho=0.2$. 
\begin{flushleft}
\textbf{Scenario 2:} Similar proportions of low-income Black and high-income White residents with moderate spatial correlation. A multivariate Gaussian distribution is used to generate the logit transformation of the proportions with means of -1.72 $(p_1=p_2=0.15)$, with a variance of 0.2. Spatial correlation is based on a proper CAR model with $\rho=0.65$. 
\end{flushleft}
\begin{flushleft}
\textbf{Case 2}\\
\textbf{High Racialized Economic Segregation (of the deprived group) with spatial variation:} We assume that the true high-income White proportions are less than the low-income Black proportions in a region.
\end{flushleft}
\textbf{Scenario 3:} Less high-income White vs. more low-income Black residents with low spatial correlation. A multivariate Gaussian distribution is used to generate the logit transformation of the proportions with means of -1.72 $(p_1=0.15)$, and -0.4 $(p_2=0.4)$ for high-income White and low-income Black residents, respectively, with a variance of 0.4. Spatial correlation is based on a proper CAR model with $\rho=0.2$.
\begin{flushleft}
\textbf{Scenario 4:} Less high-income White vs. more low-income Black residents with moderate spatial correlation.
A multivariate Gaussian distribution is used to generate the logit transformation of the proportions with means of -1.72 $(p_1=0.15)$, and -0.4 $(p_2=0.4)$ for high-income White and low-income Black residents, respectively, with a variance of 0.4. Spatial correlation is based on a proper CAR model with $\rho=0.65$.
\end{flushleft}

To see the estimates of the ICE measure for data with different population sizes, data are generated under the above scenarios for values of N of 150, 500, and 2000. 
All models are fitted with R-NIMBLE packages \citep{de2017programming}.

\section{Simulation results}
We use the root mean square error (RMSE), coverage probability, credible interval width, and WAIC values to evaluate the performance of models in various simulation settings. Specifically, the RMSE is used to assess the accuracy of models, while the coverage probability and credible interval width are used to quantify the adequacy of $95\%$ intervals. All models are compared using Watanabe-Akaike information criterion (WAIC), a  more fully Bayesian approach for estimating the out-of-sample expectation. The simulation results are shown in Table 1.1-1.3, with each table displaying the performance of three proposed approaches (6 models total) with small, medium, and large population sizes assumed. Model 1 (M1) is the bootstrap method and thus only displays the RMSE, coverage probability, and interval width. However, M1 consistently performs poorly in various simulation scenarios.  The RMSE values based on M1 are always greater than global smooth models (M2-M4) and locally smooth models (M5-M6), even though the RMSE values decrease as the sample size increases; however, it remains higher when compared to other models. When analyzing data with low racialized economic segregation (scenarios 1 and 2), M1 has consistently lower RMSE values than analyzing high racialized economic segregation data (scenarios 3 and 4). This may be due to the fact that M1 is unable to capture the sudden change between geographic space present in the simulated dataset. Bayesian spatial models, including both globally (M2-M4) and locally (M5-M6) smooth models, on the other hand, always have lower RMSE than M1, and they even have lower RMSE when analyzing highly segregated data. This is because the spatial models borrow strength from neighborhood areas and account for the spatial correlation in the estimation. The coverage probability of M1 across several scenarios is close to the nominal coverage levels of $95\%$; nevertheless, the interval widths are notably wider than M2-M4 and M5-M6, with intervals nearly twice as wide. This is because Bayesian spatial models incorporate additional random effects that account for spatial correlation and uncertainty in the study area, hence reducing the variability when estimating data. 

Models 2 through 4 are globally smooth models in which the random effects are assigned different spatial priors, including ICAR, BYM, and Leroux. Overall, the BYM model has a lower RMSE than ICAR and Leroux, which may be because it has an unstructured term that captures additional data variability in the estimation. The coverage probability of these three spatial priors is similar when the data has low segregation with a small population size $(N_i=150)$. It improves as the population size increases, as the coverage probabilities are close to the nominal coverage levels of $95\%$. When the data is highly segregated, we observe that all three priors have converge probability that are close to $95\%$, even with smaller population sizes assumed. Furthermore, the interval widths of the three global smooth models are very similar, with the Leroux model having a slightly narrower credible interval in some scenarios. In terms of model fitting, the BYM has the lowest WAIC values across all scenarios when compared to ICAR and Leroux.

The locally smooth models (M5-M6), on the other hand, perform consistently well across all scenarios. The RMSE is lower when compared to M1 and global smooth models (M2-M4), which could be because the locally smooth models we propose here can capture not only spatial correlation and uncertainty but also neighborhood areas with very different proportions. The global smooth models have the advantage of spatially smoothing the proportion surface; however, when the data is segregated or there are spatial discontinuities in the data, the global smooth models may perform poorly compared to locally smooth models.
The coverage probabilities of M5-M6 are close to the nominal levels when data are highly segregated, ranging between $89\%-94\%$; however, the coverage probabilities decrease when data are assumed to have low segregation and small population sizes. This may be because the locally smooth models attempt to model spatial correlation when there is no sufficient data to support it. When data is highly segregated, we observe that the coverage probabilities of M5-M6 increase accordingly. Although M5-M6 do not have the narrowest credible interval width across all scenarios, they are quite close to the global smooth models (M2-4) and always narrower than the bootstrap method (M1). Finally, we find that the locally smooth models (M5-M6) have lower WAIC values than the global smooth models (M2-M4) across all scenarios, implying that the locally smooth models fit the data more effectively. However, when the population size is large ($N_i=2000$), the difference in WAIC values between locally smooth models with two and three clusters becomes less significant. This remains a potential challenge for selecting the  appropriate  cluster size for the model, however, when the population size is relatively small ($N_i=150$ and $N_i=500$), we notice that their WAIC values are quite different, with the locally smooth models with three clusters having the smallest WAIC values. 

Based on the simulation results, the locally smooth model with three clusters performs well in most of the scenarios, and it will be used to construct the new ICE metric as well as its $95\%$ credible interval. 
All proposed models will be applied to the real data to determine if the results are consistent with the simulation studies.

\section{Real data analysis}

We now present the results of each proposed model based on the ICE data from Georgia at two non-overlapping time points (2005-2009 and 2016-2020). Table 2 presents the findings of the estimated ICE metric from all models, including posterior medians and $95\%$ uncertainty intervals. In both 2009 and 2020 data, M1 consistently produces slightly different (higher) estimates of ICE and wider uncertainty intervals than other models. These are also observed in the simulation studies, in which M1 has larger RMSE values and wider confidence intervals. On the other hand, the globally smooth models (M2-M4) and locally smooth models (M5-M6) have similar estimates of ICE, with the Leroux model providing a slightly narrower uncertainty interval; however, the difference of interval width between Leroux and locally smooth models is very small (less than 0.003). Table 2 also shows the WAIC values for each model, which are used as model selection criteria, with a lower WAIC value indicating better model fit. We find that the locally smooth model with three clusters ($q=3$) has the lowest WAIC values for both 2009 and 2020 data, which is consistent with simulation studies. Based on these results, we interpret our findings using the locally smooth model with $q=3$. 

\begin{figure}[ht]
   \centering 
 \includegraphics[height=10cm, width=16.5cm]{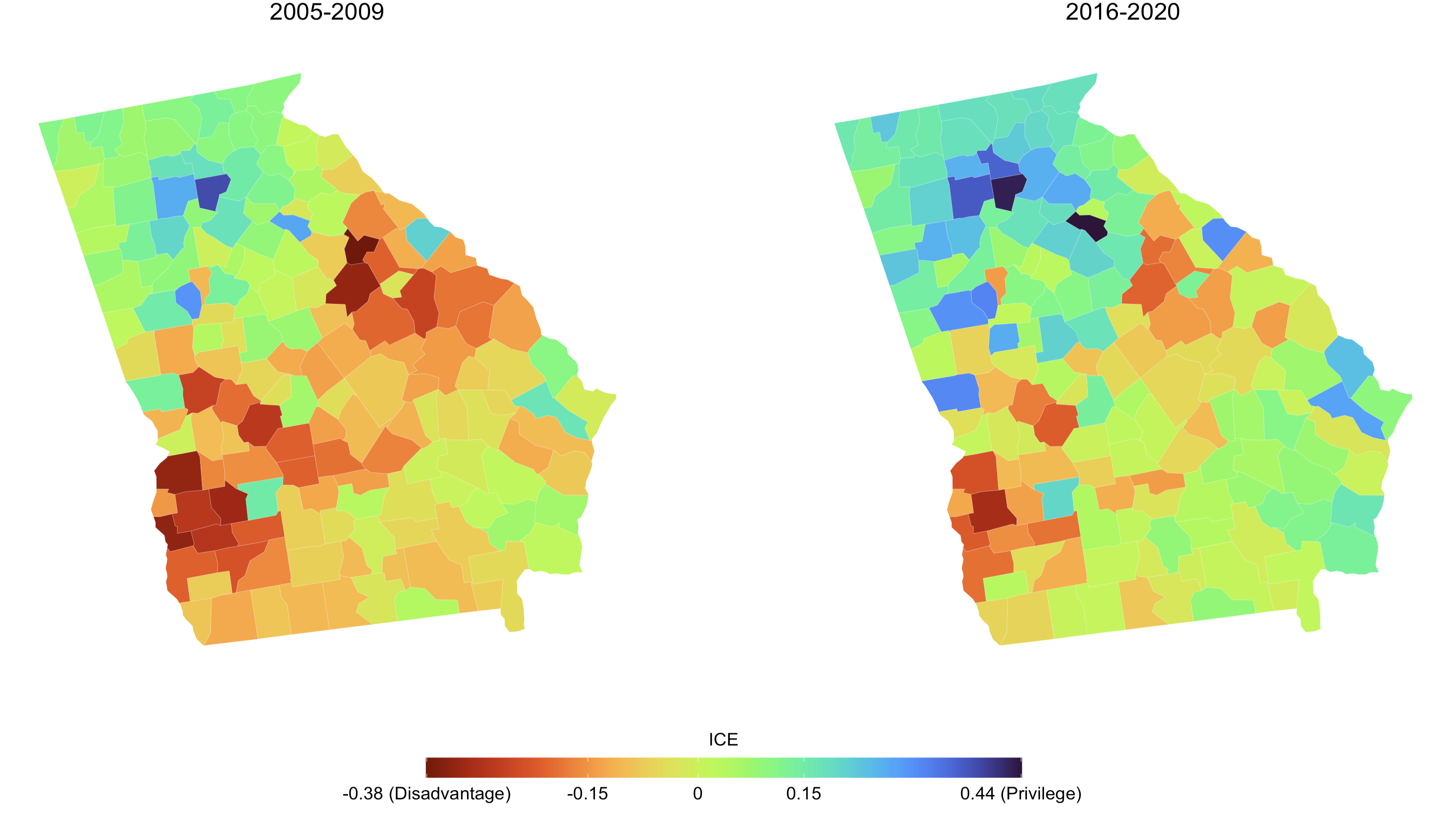}
   \caption{Choropleth maps of the posterior estimates of ICE for both 2005-2009 (left) and 2016-2020 (right) at the Georgia county level (based on locally smooth model with three clusters - M6). Darker red indicates that residents are concentrated in the disadvantaged group, and blue indicates that residents are concentrated in the privileged group.}
   \label{fig:p1}
\end{figure}

Based on the locally smooth model with $q=3$, racialized economic segregation in Georgia has increased between 2009 and 2020, as the posterior estimates of ICE are -0.04 (CrI: -0.32, 0.21) in 2009 and 0.04 (-0.21, 0.31) in 2020; however, since their $95\%$ credible intervals overlap, we conclude that this change is not statistically significant. Despite the fact that the result is not statistically significant, it is important to note that the average ICE value for Georgia has shifted sign from negative (disadvantaged low-income Black) to positive (privilege high-income White). Figure \ref{fig:p1} displays choropleth maps of ICE for both 2009 and 2020 at the Georgia county level, where darker red indicates that residents of counties are concentrated in the disadvantaged group, and blue indicates that residents of counties are concentrated in the privileged group. These maps not only reveal the spatial patterning of the ICE, but also the clustering effect in the study area. In both 2009 and 2020 data, the ICE values in the northern and southeast regions of Georgia are consistently positive, indicating that the majority of residents, on average, in these counties are Whites with high incomes. In contrast, the southwest and northeast regions of Georgia have negative ICE values, suggesting that most residents, on average, in these counties are low-income Black. Figure 2 further shows the change in ICE values in 2009 and 2020, where some counties have switched signs from negative to positive, which might suggest that: (1) increased proportion of high-income White residents and/or (2) decreased proportion of low-income Black residents. 
In addition, Figure 3 provides a clear view of which Georgia counties have changed their signs during the two non-overlapping periods. The figure shows that the majority of the sign-changed counties are in the central and southern parts of Georgia, and these counties are adjacent to each other, further demonstrating the presence of spatial and cluster patterns in the data. 
Figure 4 displays the magnitude of proportion groups (high-income White vs. low-income Black) among counties with changed signs. Counties with an increased proportion of high-income White residents and decreased proportion of low-income Black residents between two-time points are illustrated in yellow; counties with increased high-income White residents and low-income Black residents (the proportion of high-income White residents is greater than the proportion of low-income Black residents) are shown in blue. One county in red indicates that its ICE value has changed from positive to negative (Irwin county); both proportion groups have increased over time; however, the proportion of low-income Black is greater than that of high-income White.
\vspace{5mm}

\vspace{5mm}
\begin{figure}[ht]
   \centering 
   \includegraphics[height=10cm, width=15cm]{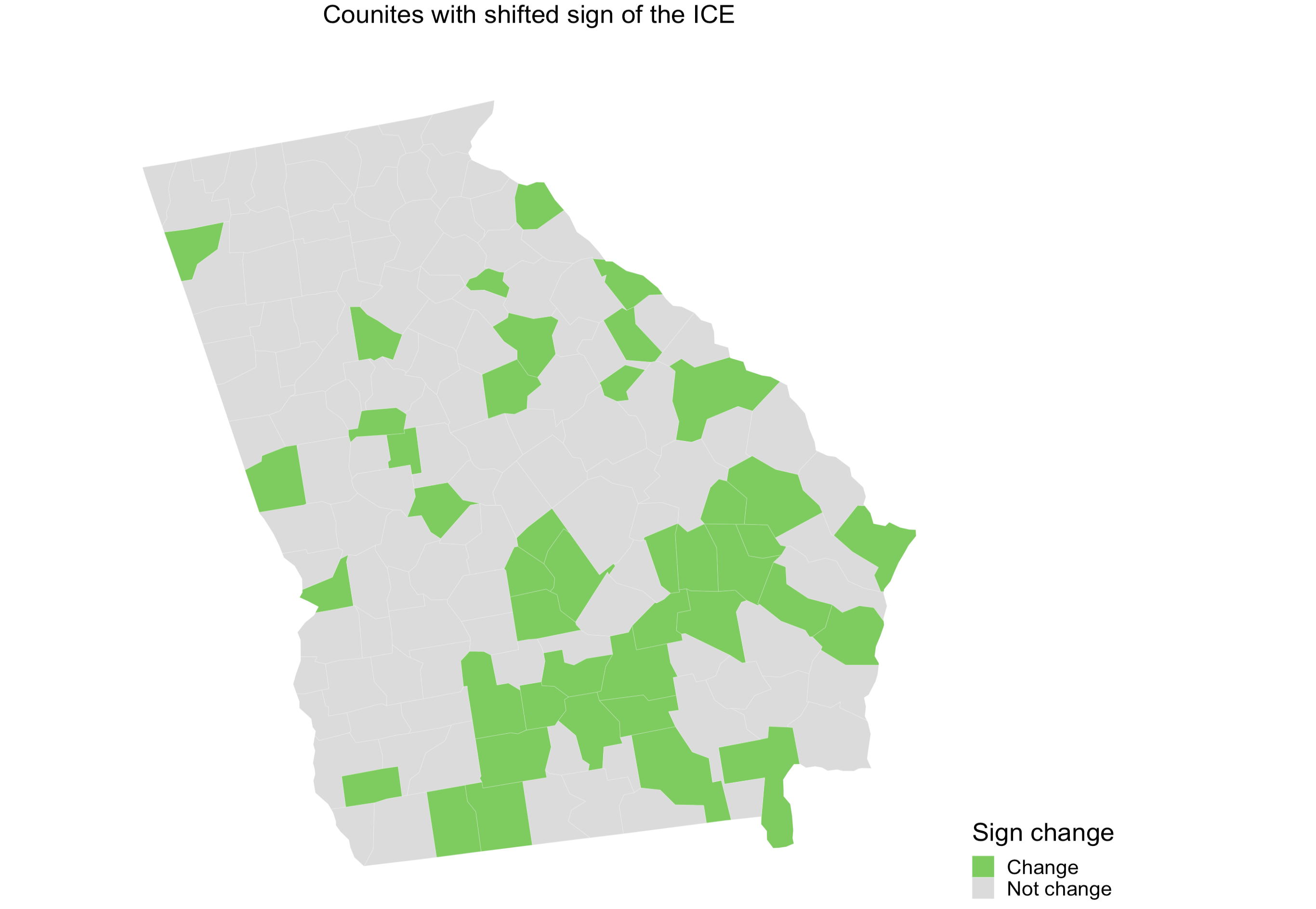}
   \caption{Map of counties with sign changed between 2005-2009 and 2016-2020.}
   \label{fig:p3}
\end{figure}

Lastly, we conduct a sensitivity analysis for each spatial model by changing the Inverse-Gamma (a,b) prior for the random effects. We use the following priors: (1) Inverse-Gamma (a=0.01, b=0.01), (2) Inverse-Gamma (a=0.1, b=0.1), and (3) Inverse-Gamma (a=0.5, b=0.0005). Our results show that the posterior estimates of ICE and it's $95\%$ credible intervals change by less than 0.0005 (see Appendix tables S1-3).



\vspace{5mm}
\begin{figure}[ht]
   \centering 
  \includegraphics[height=9cm, width=16.5cm]{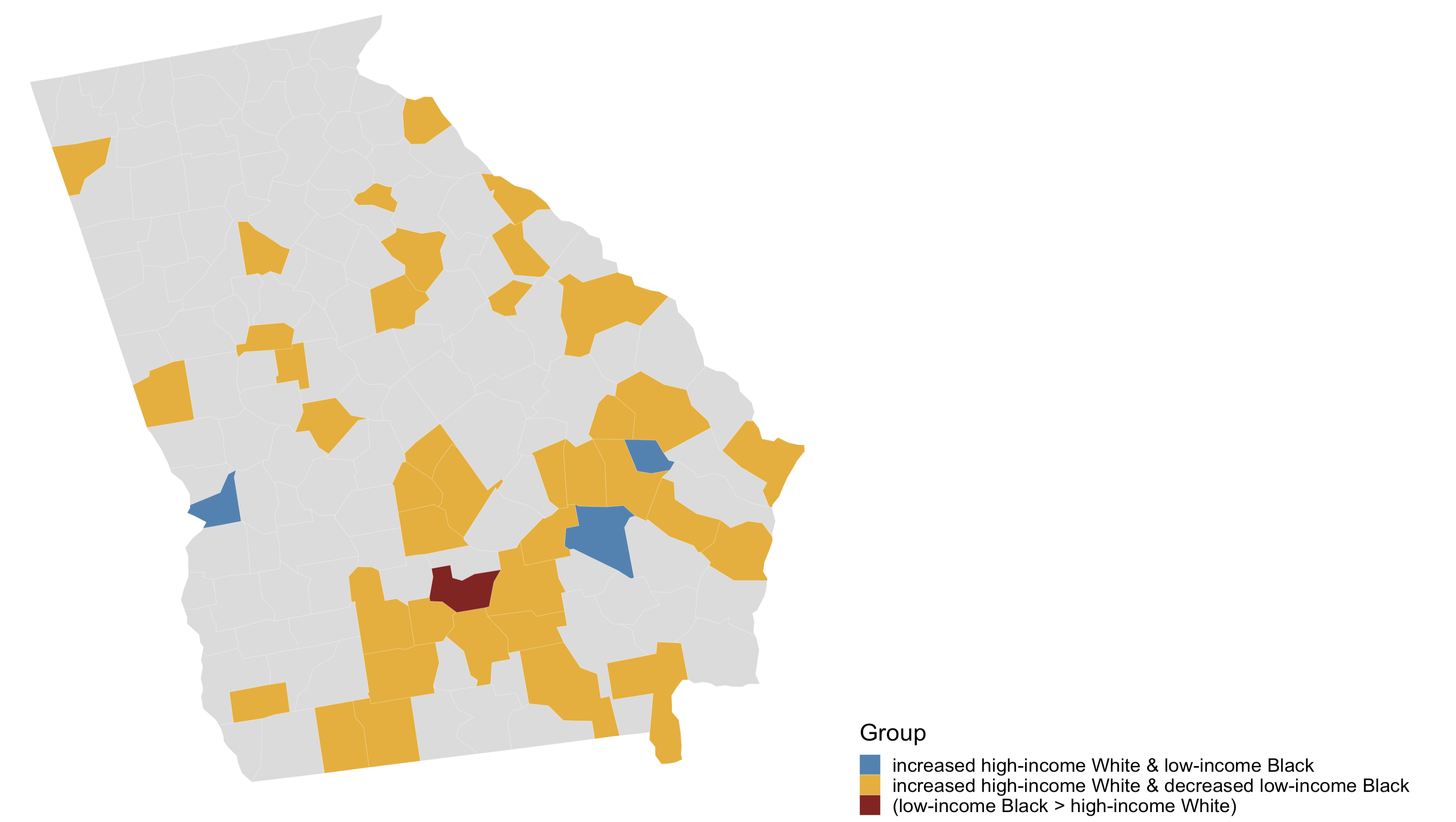}
  \caption{Map of the magnitude of proportion groups (high-income White vs. low-income Black) among counties with changed signs between 2005-2009 and 2016-2020.}
  \label{fig:p4}
  \end{figure}

\section{Discussion}

This paper describes and proposes a reformulation of the Index of Concentration at the Extremes. This formulation could account for spatial correlation and uncertainty that are present in the data when estimating the proportion groups on which the new ICE metric is based. Although the ICE metric has been extensively studied and applied in various fields, to the best of our knowledge, this is the first paper that  proposes a novel way of considering the ICE metric, notably, as a function of two sample proportions. We propose three methods for estimating the new ICE metric: bootstrap resampling, Bayesian global, and locally smooth models. Furthermore, building on a previous study (see \citealt{lee2015bayesian}), we propose an alternative version of locally smooth models which could allow researchers to implement it in other software packages (i.e., NIMBLE). These approaches are evaluated using simulation studies under various ICE scenarios. We further estimate the ICE metric based on the motivated study - racialized economic segregation at the county level in Georgia, in order to determine whether the ICE metric changed significantly between two overlapping time periods. 

Measuring segregation is intrinsically challenging, and we acknowledge that existing research emphasizes the importance of adjusting and estimating different segregation indices. For example, \citet{morrill1991measure} and \citet{wong2003spatial} proposed to adjust the Dissimilarity index to account for the spatial correlation in the data, whereas \citet{lee2015bayesian} proposed using the standard formulation of the Dissimilarity index by estimating the proportion surface in the formulation to account for the uncertainty and spatial correlation.  Recently, \citet{hu2022bayesian} proposed a mixture of finite mixture models to capture the spatial homogeneity of Lorenz curves, a function that shows the proportion of total income received by the bottom $100p\%$ $(p\in [0,1])$  of the population and also derives the Gini coefficient, a popular indicator of income inequality. In addition to the aforementioned research, our paper is one of the few that examines the formulation of the ICE metric and addresses the importance of the nature of the data used to compute it. 

On the basis of the simulation study, we demonstrate that the bootstrap method (M1) is not appropriate for recomputing the ICE metric, as it consistently yields larger RMSE values and wider confidence intervals under different residential segregation scenarios. Bayesian globally smooth models (M2-M4), on the other hand, outperform M1 with lower RMSE values and narrower credible intervals since they are able to account for spatial correlation and uncertainty in the data. The simulation study further has shown that the proposed locally smooth models consistently perform well across all scenarios, with the lowest RMSE values. Its coverage probability improved as sample sizes increased in a region, and its credible intervals are comparable to those of globally smooth models. Lastly, based on the WAIC values, the locally smooth model has the lowest WAIC values across all scenarios, which suggest that the proposed model shows good model fit and efficiency. The results from the simulation indicate that the locally smooth model is appropriate for constructing the new ICE metric, as it not only accounts for spatial correlation and uncertainty in the estimation, but also captures the cluster effect that may be present in the data. 

The results from our motivated study show that racialized economic segregation in Georgia has shifted between 2005-2009 and 2016-2020, with the ICE metric increasing from -0.04 (2005-2009) to 0.04 (2016-2020), although the $95\%$ credible intervals overlap. Figure 3 (b) shows details regarding this change; that is, the majority of sign-changed counties have increased high-income White proportions and decreased low-income Black proportions. Although the causes of this change in Georgia may be due to a combination of multiple factors, our findings will be helpful for those who consider racialized economic segregation as a social determinant of health and who are interested in assessing the relationship between the change in racialized economic segregation and various health outcomes in areal units. 

The generality of our proposed study can be seen in the following aspects. Firstly, we argue that the data used to construct the ICE metric are from survey data and are spatial in nature; since this metric includes both race/ethnicity and income segregation in area units, spatial correlation and uncertainty should be included in computing this metric, which they are not currently. The existing formulation for computing the ICE metric makes it challenging to address the aforementioned issues; therefore, we proposed reformulating the ICE metric by addressing on its essential components - a function of two sample proportions. Secondly, the new formulation of the ICE metric allows for estimating each proportion group and accounting for the spatial correlation present in the data, as demonstrated by the racialized economic segregation data in Georgia, where each proportion group (high-income White and low-income Black) shows significant Moran's I statistics, indicating the presence of spatial correlation. Thirdly, we have demonstrated that the proposed locally smooth model performs well in simulation studies and real data analysis, and that it can be implemented with greater flexibility in other software packages. Finally, although we have used modeling approaches to objectively compare the change of racialized economic segregation at two non-overlapping time points in Georgia, it may be worthwhile for future research to develop a spatio-temporal model to evaluate the effect of the ICE metric on specific health outcomes at multiple time points.

\newpage
\bibliographystyle{jasa}        
\bibliography{ref}

\clearpage

\setcounter{table}{0}
\renewcommand\thetable{1.1}
{\renewcommand{\arraystretch}{1.5}
\begin{table}[ht]
\centering
\vspace{-1.2cm}
\hspace*{-1.3cm}
\begin{tabular}{ccccccc}
\toprule
\multicolumn{1}{c}{\textbf{N=150}} \\
\hline
\textbf{ } & \textbf{M1-Bootstrap} & \textbf{M2-BYM} & \textbf{M3-ICAR} & \textbf{M4-Leroux} &\textbf{M5-L2}& \textbf{M6-L3}  \\
\midrule
RMSE  &   \\
1  & 0.0568 & 0.0321 & 0.0392 & 0.0425 & 0.0311 & 0.0304\\
2 & 0.0582 & 0.0315 & 0.0368 & 0.0387 & 0.0305 & 0.0298\\
3 & 0.2139 & 0.0233 & 0.0250 & 0.0276 & 0.0228 & 0.0225\\
4 & 0.2151 & 0.0232 & 0.0250 & 0.0255 & 0.0227 & 0.0223 \\
\hline
Coverage &   \\
1 & 95.3 & 60.6 & 53.9 & 42.8 &  62.2 & 63.1 \\
2 & 95.3 & 64.7 & 60.4 & 52.9 & 65.7 & 66.2 \\
3 & 1 & 89.4 & 89.7 & 89.5 & 89.5 & 88.8\\
4 & 1 & 89.6 & 89.2 & 88.6 & 89.4 & 89.0\\
\hline
Interval width\\
1 & 0.2272 & 0.0976 & 0.0843 & 0.0652 & 0.1000 & 0.1025\\
2 & 0.2328 & 0.1092 & 0.0994 & 0.0851 & 0.1111 & 0.1130\\
3 & 0.8566 & 0.6615 & 0.6599 & 0.6586 & 0.6618 & 0.6571\\
4 & 0.8607 & 0.6702 & 0.6601 & 0.6543 & 0.6637 & 0.6675\\
\hline
WAIC\\
1 &--- & 2013.866 & 2080.246 & 2145.317 & 2006.678 & 2002.947\\
2 &--- & 2018.155 & 2052.559 & 2218.705 & 2014.178 & 2011.213\\
3 &--- & 2136.558 & 2154.918 & 2809.597 & 2139.792 & 2132.231\\
4 &--- & 2136.830 & 2136.081 & 2327.886 & 2136.728 & 2132.216\\

\bottomrule
\end{tabular}\hspace*{-1.3cm}\captionsetup{width=1\textwidth}
\caption{Results of simulation study with small sample size (N=150) for all models.
1= Low Racialized Economic Segregation with little spatial variation; 2= Low Racialized Economic Segregation with moderate spatial variation; 3=High Racialized Economic Segregation (of the deprived group) with little spatial variation; and 4=High Racialized Economic Segregation (of the deprived group) with moderate spatial variation.}
  \label{tab:t2}
\end{table}}

\clearpage
\renewcommand\thetable{1.2}
{\renewcommand{\arraystretch}{1.5}
\begin{table}[ht]
\centering
\vspace{-1cm}
\hspace*{-1.3cm}
\begin{tabular}{ccccccc}
\toprule
\multicolumn{1}{c}{\textbf{N=500}} \\
\hline
\textbf{ } & \textbf{M1-Bootstrap} & \textbf{M2-BYM} & \textbf{M3-ICAR} & \textbf{M4-Leroux} &\textbf{M5-L2}& \textbf{M6-L3}  \\
\midrule
RMSE  &   \\
1  & 0.0352 & 0.0154 & 0.0181 & 0.0172 & 0.0147 & 0.0144  \\
2  & 0.0371 & 0.0154 & 0.0173 & 0.0169 & 0.0146 & 0.0143\\
3  & 0.2055 & 0.0084 & 0.0091 & 0.0091 & 0.0083 & 0.0089 \\
4  & 0.2067 & 0.0085 & 0.0091 & 0.0096 & 0.0083 & 0.0086 \\
\hline
Coverage &   \\
1  & 95.4 & 74.5 & 70.1 & 69.8 & 73.6 & 75.7 \\
2  & 95.2 & 76.3 & 74.0 & 72.7 & 76.3 & 76.9\\
3  & 1 & 92.1 & 91.9 & 92.3 & 91.8  & 91.5\\
4  & 1 & 92.3 & 93.2 & 92.2 & 92.1 & 92.1\\
\hline
Interval width\\
1 & 0.1401 & 0.0806 & 0.0749 & 0.0750 & 0.0804 & 0.0824\\
2 & 0.1485 & 0.0873 & 0.0842 & 0.0834 & 0.0869 & 0.0884\\
3 & 0.8221 & 0.6274 & 0.6269 & 0.6288 & 0.6249 & 0.6236\\
4 & 0.8270 & 0.6384 & 0.6376 & 0.6362 & 0.6329 & 0.6365\\
\hline
WAIC\\
1 &--- & 2414.568 & 2487.727 & 2651.186 & 2411.628 & 2404.867\\
2 &--- & 2407.207 & 2447.625 & 2729.141 & 2406.874 & 2403.697 \\
3 &--- & 2530.401 & 2534.217 & 3657.600 & 2534.557 & 2525.431\\
4 &--- & 2525.054 & 2536.481 & 3586.350 & 2532.795 & 2518.486\\

\bottomrule
\end{tabular}\hspace*{-1.3cm}\captionsetup{width=.99\textwidth}
\caption{Results of simulation study with middle sample size (N=500) for all models.
1= Low Racialized Economic Segregation with little spatial variation; 2= Low Racialized Economic Segregation with moderate spatial variation; 3=High Racialized Economic Segregation (of the deprived group) with little spatial variation; and 4=High Racialized Economic Segregation (of the deprived group) with moderate spatial variation.}
  \label{tab:t3}
\end{table}}

\clearpage
\renewcommand\thetable{1.3}
{\renewcommand{\arraystretch}{1.5}
\begin{table}[ht]
\centering
\vspace{-1cm}
\hspace*{-1.3cm}
\begin{tabular}{ccccccc}
\toprule
\multicolumn{1}{c}{\textbf{N=2000}} \\
\hline
\textbf{ } & \textbf{M1-Bootstrap} & \textbf{M2-BYM} & \textbf{M3-ICAR} & \textbf{M4-Leroux} &\textbf{M5-L2}& \textbf{M6-L3}  \\
\midrule
RMSE  &   \\
1  & 0.0223 & 0.0054 & 0.0071 & 0.0069 & 0.0058 & 0.0052  \\
2  & 0.0278 & 0.0041 & 0.0058 & 0.0059 & 0.0049 & 0.0038\\
3  & 0.1994 & 0.0025 & 0.0027 & 0.0036 & 0.0025 & 0.0024\\
4  & 0.2018 & 0.0024 & 0.0026 & 0.0036 & 0.0024 & 0.0024 \\
\hline
Coverage &   \\
1 & 95 & 88.0 & 83.4 & 83.4 & 84.7 & 83.6 \\
2 & 95.5 & 88.8 & 88.3 & 87.6 & 88.7 & 88.9\\
3 & 1 & 93.3 & 93.2 & 93.4 & 93.1 & 93.3 \\
4 & 1 & 93.5 & 93.2 & 93.3 & 93.2 & 93.6 \\
\hline
Interval width\\
1 & 0.0892 & 0.0634 & 0.0628 & 0.0630 & 0.0643 & 0.0628\\
2 & 0.1115 & 0.0907 & 0.0894 & 0.0889 & 0.0909 & 0.0908\\
3 & 0.7977 & 0.5710 & 0.5684 & 0.5690 & 0.5689 & 0.5710\\
4 & 0.8072 & 0.6005 & 0.5938 & 0.5939 & 0.5975 & 0.6005\\
\hline
WAIC\\
1 &--- & 2893.413 & 2920.041 & 3533.694 & 2888.988 & 2880.947\\
2 &--- & 2890.986 & 2910.374 & 4290.116 & 2886.967 & 2886.835\\
3 &--- & 2970.264 & 2976.110 & 3372.040 & 2970.028 & 2968.926\\
4 &--- & 2972.600 & 2973.002 & 3258.700 & 2972.023 & 2970.996\\

\bottomrule
\end{tabular}\hspace*{-1.3cm}\captionsetup{width=.99\textwidth}
\caption{Results of simulation study with large sample size (N=2000) for all models.
1= Low Racialized Economic Segregation with little spatial variation; 2= Low Racialized Economic Segregation with moderate spatial variation; 3=High Racialized Economic Segregation (of the deprived group) with little spatial variation; and 4=High Racialized Economic Segregation (of the deprived group) with moderate spatial variation.}
  \label{tab:t4}
\end{table}}

\clearpage
\renewcommand\thetable{2}
{\renewcommand{\arraystretch}{2}
\begin{table}[h]
\centering
\begin{tabular}{lcccc}
\toprule
\multicolumn{1}{c}{} & \multicolumn{2}{c}{\textbf{2005-2009 data}} & \multicolumn{2}{c}{\textbf{2016-2020 data}} \\
\cmidrule(rl){2-3} \cmidrule(rl){4-5}
\textbf{Model} & \textbf{WAIC} & \textbf{ICE} & \textbf{WAIC} & \textbf{ICE}  \\
\midrule
M1-Bootstrap   & --- & -0.0333 (-0.2988, 0.2322) & --- & 0.0504 (-0.2259, 0.3267)  \\
\hline
M2-BYM    & 3400.413 & -0.0422 (-0.3188, 0.2054) & 3666.200 & 0.0401 (-0.2084, 0.3086)  \\
M3-ICAR   & 3891.896 & -0.0422 (-0.3192, 0.2053) & 4117.865 & 0.0394 (-0.2084, 0.3086)\\
M4-Leroux & 3326.957 & -0.0416 (-0.3157, 0.2051) & 3698.617 & 0.0389 (-0.2054, 0.3089)  \\
\hline
M5-L2 & 3335.835 & -0.0421 (-0.3197, 0.2053) & 3561.076 & 0.0401 (-0.2084, 0.3086)  \\
M6-L3 & 3326.512 & -0.0421 (-0.3192, 0.2053) & 3511.743 &0.0395 (-0.2084, 0.3086) \\
\bottomrule
\end{tabular}\captionsetup{width=.99\textwidth}
\caption{Estimates and $95\%$ uncertainty intervals  of Racialized economic segregation (ICE: Index of Concentration at the Extremes). WAIC: Watanabe–Akaike information criterion. M1: model 1, bootstrap approach; M2-4: global smooth models, including Besag York Mollié Model (BYM); Intrinsic Conditional Auto-Regressive model (ICAR); Leroux model (Leroux); M5-6: locally smooth models, with two clusters (L2) and three clusters (L3).}
  \label{tab:t1}
\end{table}}

\clearpage

\renewcommand\thetable{S1}
{\renewcommand{\arraystretch}{2}
\begin{table}[h]
\centering
\begin{tabular}{lcc}
\toprule
\textbf{Model} & \textbf{2005-2009} & \textbf{2016-2020} \\
\hline
M1-Bootstrap   & --- & ---\\
\hline
M2-BYM    & -0.0422 (-0.3195, 0.2054) &0.0401 (-0.2086, 0.3086)\\

M3-ICAR  &-0.0420 (-0.3190, 0.2053)& 0.0398 (-0.2086, 0.3085)\\

M4-Leroux &-0.0424 (-0.3179, 0.2051)& 0.0400 (-0.2043, 0.3085)\\
\hline
M5-L2 & -0.0420 (-0.3192, 0.2054)& 0.0398 (-0.2080, 0.3086)  \\
M6-L3 &-0.0420 (-0.3198, 0.2054) &0.0397 (-0.2078, 0.3086) \\
\hline
\end{tabular}\captionsetup{width=.75\textwidth}
\caption{Sensitivity analysis with Inverse-Gamma (0.1, 0.1) for the random effects.}
  \label{tab:s1}
\end{table}}

\clearpage
\renewcommand\thetable{S2}
{\renewcommand{\arraystretch}{2}
\begin{table}[h]
\centering
\begin{tabular}{lcc}
\toprule
\textbf{Model} & \textbf{2005-2009} & \textbf{2016-2020} \\
\hline
M1-Bootstrap   & --- & ---\\
\hline
M2-BYM    & -0.0421 (-0.3190, 0.2055) & 0.0399 (-0.2084, 0.3086)\\

M3-ICAR  &-0.0420 (-0.3188, 0.2054)&  0.3963 (-0.2078, 0.3085)\\

M4-Leroux &-0.0388 (-0.3184, 0.2053) & 0.0415 (-0.2126, 0.3084)\\
\hline
M5-L2 & -0.0421 (-0.3190, 0.2053)& 0.0400 (-0.2084, 0.3085) \\
M6-L3 &-0.0420 (-0.3194, 0.2053) & 0.0400 (-0.2080, 0.3086)\\
\hline
\end{tabular}\captionsetup{width=.75\textwidth}
\caption{Sensitivity analysis with Inverse-Gamma (0.01, 0.01) for the random effects.}
  \label{tab:s2}
\end{table}}

\clearpage
\renewcommand\thetable{S3}
{\renewcommand{\arraystretch}{2}
\begin{table}[h]
\centering
\begin{tabular}{lcc}
\toprule
\textbf{Model} & \textbf{2005-2009} & \textbf{2016-2020} \\
\hline
M1-Bootstrap   & --- & ---\\
\hline
M2-BYM    & -0.0421 (-0.3189, 0.2054)& 0.0398 (-0.2079, 0.3087)\\

M3-ICAR  &-0.0420 (-0.3188, 0.2053) &	0.0401 (-0.2086, 0.3086)\\

M4-Leroux &-0.0419 (-0.3206, 0.2050)& 0.0360 (-0.2083, 0.3085)\\
\hline
M5-L2 & -0.0421 (-0.3190, 0.2054) & 0.0398 (-0.2085, 0.3087)\\
M6-L3 &-0.0421 (-0.3194, 0.2054)& 0.0397 (-0.2085, 0.3087)\\
\hline
\end{tabular}\captionsetup{width=.75\textwidth}
\caption{Sensitivity analysis with Inverse-Gamma (0.5, 0.0005) for the random effects.}
  \label{tab:s3}
\end{table}}

\clearpage

\renewcommand\thetable{S4}
{\renewcommand{\arraystretch}{1.2}
\begin{table}[ht]
\centering
\begin{tabular}{l|cc}
\toprule
\textbf{County with shifted ICE} & \textbf{Fips code} & \textbf{County Name} \\
\hline
From negative to positive  & 13001 & Appling County  \\
     & 13003 & Atkinson County\\
     & 13019 & Berrien County\\
     &13023 &	Bleckley County\\
     &13031	&Bulloch County\\
&13033	&Burke County\\
&13043	&Candler County\\
&13049	&Charlton County\\
&13051	 &Chatham county\\
&13053	&Chattahoochee County\\
&13055	&Chattooga County\\
&13059	&Clarke County\\
&13065	&Clinch County\\
&13069	 &Coffee County\\
&13071	&Colquitt County\\
&13079	&Crawford County\\
&13089	&DeKalb County\\
&13091	&Dodge County\\
&13109	&Evans County\\
&13125	&Glascock County\\
&13131	&Grady county\\
&13133	&Greene County\\
&13147	&Hart County\\
&13161	&Jeff Davis County\\
&13171	 &Lamar County\\
&13181	 &Lincoln County\\
&13183	&Long County\\
&13189	&McDuffie County\\
&13191	&McIntosh County\\
&13201	&Miller County\\
&13209	&Montgomery county\\
&13235	&Pulaski County\\
\hline
\end{tabular}\captionsetup{width=.75\textwidth}
\caption{Counties with shifted ICE measure between 2005-2009 and 2016-2020.}
  \label{tab:s4}
\end{table}}

\clearpage
\renewcommand\thetable{S4.c}
{\renewcommand{\arraystretch}{1.2}
\begin{table}[ht]
\centering
\begin{tabular}{l|cc}
\toprule
\textbf{County with shifted ICE} & \textbf{Fips code} & \textbf{County Name} \\
\hline
From negative to positive  
&13237	&Putnam County\\
&13255	 &Spalding County\\
&13267	&Tattnall County\\
&13275	&Thomas County\\
&13277	&Tift County\\
&13279	&Toombs County\\
&13285	&Troup County\\
&13315	& Wilcox County\\
&13321	&Worth County\\
\hline
From positive to negative & 13155 & Irwin County\\
\hline
\end{tabular}\captionsetup{width=.68\textwidth}
\caption{Counties with shifted ICE measure between 2005-2009 and 2016-2020.}
  \label{tab:s4.c}
\end{table}}

\end{document}